\documentclass[lettersize,journal,twoside]{IEEEtran}
\usepackage{amsmath,amsfonts}
\usepackage{algorithmic}
\usepackage{algorithm}
\usepackage{array}
\usepackage[caption=false,font=normalsize,labelfont=sf,textfont=sf]{subfig}
\usepackage{textcomp}
\usepackage{stfloats}
\usepackage{url}
\usepackage{verbatim}
\usepackage{graphicx}
\usepackage{cite}
\usepackage{gensymb}
\usepackage{orcidlink}
\hypersetup{hidelinks}
\usepackage[absolute,overlay]{textpos}

\begin{document}

\title{Multi-Mode Process Control Using Multi-Task Inverse Reinforcement Learning}

\author{Runze Lin\orcidlink{0000-0002-2874-105X},
	Junghui Chen\orcidlink{0000-0002-9994-839X},
	Biao Huang\orcidlink{0000-0001-9082-2216},~\IEEEmembership{Fellow,~IEEE},
	Lei Xie\orcidlink{0000-0002-7669-1886},
	Hongye Su\orcidlink{0000-0002-7003-1000},~\IEEEmembership{Senior Member,~IEEE}
\thanks{This work was supported in part by the National Key R\&D Program of China under Grant 2022YFB3305903, and the National Science and Technology Council, Taiwan under Grant NSTC 113-2221-E-033-003.
}
\thanks{R. Lin, L. Xie, and H. Su are with the State Key Laboratory of Industrial Control Technology, Institute of Cyber-Systems and Control, Zhejiang University, Hangzhou 310027, China (e-mail: rzlin@zju.edu.cn; leix@iipc.zju.edu.cn; hysu@iipc.zju.edu.cn).}
\thanks{J. Chen is with the Department of Chemical Engineering, Chung-Yuan Christian University, Taoyuan 32023, Taiwan, R.O.C. (e-mail: jason@wavenet.cycu.edu.tw).}
\thanks{B. Huang is with the Department of Chemical and Materials Engineering, University of Alberta, Edmonton, AB T6G 2G6, Canada (e-mail: biao.huang@ualberta.ca).}}

\markboth{IEEE Transactions on Cybernetics, April~2025}%
{Lin \MakeLowercase{\textit{et al.}}: Multi-Mode Process Control Using Multi-Task Inverse Reinforcement Learning}

\maketitle

\begin{abstract}
In the era of Industry 4.0 and smart manufacturing, process systems engineering must adapt to digital transformation. While reinforcement learning offers a model-free approach to process control, its applications are limited by the dependence on accurate digital twins and well-designed reward functions. To address these limitations, this paper introduces a novel framework that integrates inverse reinforcement learning (IRL) with multi-task learning for data-driven, multi-mode control design. Using historical closed-loop data as expert demonstrations, IRL extracts optimal reward functions and control policies. A latent-context variable is incorporated to distinguish modes, enabling the training of mode-specific controllers. Case studies on a continuous stirred tank reactor and a fed-batch bioreactor validate the effectiveness of this framework in handling multi-mode data and training adaptable controllers.
\end{abstract}


\section{Introduction}

\begin{textblock*}{20cm}(1.37cm, 0.3cm)
	\scriptsize
	This work has been submitted to the IEEE for possible publication. Copyright may be transferred without notice, after which this version may no longer be accessible.
\end{textblock*}

\IEEEPARstart{I}{n} the era of Industry 4.0 and smart manufacturing, intelligent process control systems have become increasingly pivotal. The integration of AI into science and engineering has ushered in a new paradigm, with Deep Reinforcement Learning (DRL) offering transformative potential for modern process industries. In recent years, there has been growing interest in applying DRL to process control, leading to significant advancements in both continuous and batch process applications \cite{RN27, RN100}. A key development in this area is the incorporation of transfer learning into DRL frameworks, enhancing the safety, robustness, and practical deployment of process control systems \cite{RN80, RN87, RN92, RN75, RN76, RN85}.

Current research has yet to fully address key challenges in applying DRL to process control, such as the high costs of trial-and-error learning, low sample efficiency, and exploration instability. Industrial settings have amassed vast amounts of historical closed-loop data from PLC or DCS-control operations. However, traditional control strategies like MPC, as well as advanced algorithms such as DRL, have not fully harnessed the valuable insights embedded in this industrial big data. Extracting control patterns from real-world closed-loop data could serve as a robust foundation for DRL transfer learning \cite{RN93, RN12}.

Learning from demonstrations involves various techniques for training Reinforcement Learning (RL) controllers using expert demonstrations \cite{RN20}, with imitation learning being a widely used approach. In industrial production, abundant historical closed-loop data can be utilized to derive controller priors through imitation learning \cite{RN16}. These RL controllers can then be fine-tuned via transfer learning in real-world processes, improving the safety of DRL training. A basic method, behavior cloning, fits ``state $x_t$-action $u_t$'' pairs from expert trajectories. In contrast, Inverse RL (IRL), particularly Adversarial IRL (AIRL) \cite{RN11}, offers a more sophisticated approach by framing learning from demonstrations as a probabilistic inference problem \cite{RN10}. Nevertheless, conventional IRL methods struggle to address the multi-mode nature of process control, where varying operating modes lead to distinct data distributions, complicating their application in multi-mode controller design.

Operating processes across different modes is far from ideal, yet it remains a common and realistic challenge that poses significant difficulties for DRL. PSE researchers are well-acquainted with multi-mode processes, particularly in fields such as data-driven process monitoring, soft sensing, and fault diagnosis, where multi-mode modeling is frequently required \cite{RN7}. This necessitates the development of models tailored to distinct data distributions. Common approaches include Kernel Principal Component Analysis, Kernel Partial Least Squares, Gaussian Mixture Models, Variational Autoencoders, and InfoGAN \cite{RN9}. These methods typically utilize latent variables to capture the variability across operating modes and data distributions in the modeling process.

However, while these methods have advanced modeling capabilities, they fall short in addressing the complexities of optimizing and controlling multi-mode systems. Robust control, though capable of managing variability, is often overly conservative. Approaches like controller fusion, weighting or gain scheduling show promise but require manual design for each mode and tailored integration strategies. These methods also struggle with generalization, particularly when confronted with extrapolated conditions or unforeseen modes. Additionally, MPC relies on predefined models to capture process dynamics, but in multi-mode scenarios, model-plant mismatches frequently arise, leading to suboptimal control performance.

Although different modes involve distinct characteristics, many share common features since only certain operating conditions or materials vary. This observation leads to an important question: \textit{Can concepts from multi-mode modeling be leveraged to design a multi-mode controller using latent variables?} Given that IRL offers a robust probabilistic inference framework for offline, data-driven control design \cite{RN93}, this paper proposes augmenting IRL with latent variables and multi-task learning. The approach enables the learning of multi-mode controller priors from historical data across different modes, creating a universal controller architecture within the IRL framework. This facilitates multi-mode adaptability, as multi-task learning integrates prior knowledge from various modes \cite{RN14}. Therefore, the multi-mode controller can quickly adapt to new, unseen scenarios.

This paper introduces a novel framework for developing a multi-mode process controller using multi-task IRL. The proposed method incorporates adversarial reward learning, variational inference, and MaxEnt IRL to train a purely data-driven controller capable that can adapt to various operating modes. To the best of our knowledge, this is the first study to apply multi-task IRL for designing multi-mode process controllers in a data-driven context. This approach offers several advantages: 1) reducing the safety risks associated with DRL’s direct interaction with the environment, 2) fully utilizing the latent controller features embedded in historical industrial data, and 3) addressing the complexities of multi-mode control design through multi-task learning.

The remainder of this paper is structured as follows: Section \ref{Preliminaries} reviews the preliminaries, including RL, IRL, and AIRL. Section \ref{Problem_statement} outlines the motivation and problem statement. In Section \ref{Methodology}, the proposed methodology is presented. It details the multi-task IRL approach for addressing the multi-mode process control problem. Section \ref{Experiment} covers the experimental setup and result analysis. Finally, Section \ref{Conclusion} concludes by summarizing the key contributions of this work.

\section{Preliminaries}\label{Preliminaries}
\subsection{Markov decision process (MDP) and RL}\noindent
The MDP in RL is defined by a tuple $({\cal X},{\cal U},{p_{\cal X}},\eta ,r,\gamma )$, where ${\cal X}$ and ${\cal U}$ represent the state and action spaces respectively, ${p_{\cal X}}:{\cal X} \times {\cal U} \times {\cal X} \to [0,1]$ denotes the state transition probability of the next state ${x_{t + 1}} \in {\cal X}$ given the current state ${x_t} \in {\cal X}$ and action ${u_t} \in {\cal U}$, $\eta :{\cal X} \to {\cal P}({\cal X})$ is the initial state distribution, $r:\mathcal{X}\times \mathcal{U}\to \mathbb{R}$ specifies the reward function ${r_t} \buildrel \Delta \over = r({x_t},{u_t})$, and $\gamma  \in (0,1]$ is the discount factor. Let $\pi $ denote the control policy that selects the optimal action ${u_t} \sim \pi ({u_t}|{x_t})$ based on the current state ${x_t}$, and the state-action marginal distribution induced by the policy is represented as ${\rho _\pi }({x_t},{u_t})$.

The objective of RL is to learn a control policy that maximizes long-term (discounted) rewards through interactions with the environment $E$ in a trial-and-error manner. Formally, the standard RL problem is defined as follows:
\begin{equation}
{{\pi }^{*}}=\arg \underset{\pi }{\mathop{\max }}\,\sum\limits_{t=1}^{T}{{{\mathbb{E}}_{({{x}_{t}},{{u}_{t}})\sim{{\rho }_{\pi }}}}}[r({{x}_{t}},{{u}_{t}})]\label{eq:1}
\end{equation}

\subsection{Maximum entropy RL}\noindent
The maximum entropy (MaxEnt) RL objective is defined as:
\begin{equation}
{{\pi }^{*}}=\arg \underset{\pi }{\mathop{\max }}\,\sum\limits_{t=1}^{T}{{{\mathbb{E}}_{({{x}_{t}},{{u}_{t}})\sim{{\rho }_{\pi }}}}}[r({{x}_{t}},{{u}_{t}})+\mathcal{H}(\pi ({{u}_{t}}|{{x}_{t}}))]\label{eq:2}
\end{equation}
where $\mathcal{H}(\pi )={{\mathbb{E}}_{\pi }}[-\log \pi (u|x)]$ is the entropy-regularization term for the control policy. Unlike traditional RL, which focuses solely on maximizing the expected sum of rewards, MaxEnt RL aims to develop a policy that maximizes the likelihood of exploring favorable conditions to achieve the primary objective outlined in Eq. \eqref{eq:1}. This approach allows the learned policy to account for sub-optimal or stochastic behaviors that conventional RL may overlook, thereby providing valuable insights for inverse RL and enhancing transfer learning through exploration.

In addition to Eq. \eqref{eq:2}, there is another explanation and derivation of MaxEnt RL. Define the trajectory $\tau  \buildrel \Delta \over = \{ {x_{1:T}},{u_{1:T}}\} $ as a sequence of state-action pairs generated by a specific control policy $\pi ({u_t}|{x_t})$. The trajectory distribution under this policy $\pi$ can be expressed as follows:
\begin{equation}
{p_\pi }(\tau ) = \eta ({x_1})\prod\limits_{t = 1}^T p ({x_{t + 1}}|{x_t},{u_t})\pi ({u_t}|{x_t})\label{eq:3}
\end{equation}
In the context of probabilistic graphical model, MaxEnt RL reframes the RL control problem as an inference problem that can be addressed using the probability theory. Assuming the existence of a ground-truth reward function $r({x_t},{u_t})$, the objective of MaxEnt RL is to learn a policy from the following optimal trajectory distribution:
\begin{equation}
\begin{array}{c}
p(\tau ) \buildrel \Delta \over = \frac{1}{Z}\eta ({x_1})\prod\limits_{t = 1}^T {p({x_{t + 1}}|{x_t},{u_t})} \exp (r({x_t},{u_t}))\\
\propto \left[ {\eta ({x_1})\prod\limits_{t = 1}^T p ({x_{t + 1}}|{x_t},{u_t})} \right]\exp \left( {\sum\limits_{t = 1}^T r ({x_t},{u_t})} \right)
\end{array}\label{eq:4}
\end{equation}
where the partition function $Z = \int {\eta ({x_1})\prod\nolimits_t {p({x_{t + 1}}|{x_t},{u_t})} \exp (r({x_t},{u_t}))d\tau }$ acts as a normalizer enforcing $p(\tau ) \in [0,1]$ and $\int {p(\tau )d\tau }  = 1$. Now let us consider the KL divergence between ${p_\pi }(\tau )$ and $p(\tau )$ as follows:
\begin{equation}
\begin{aligned}
& -{{D}_{\text{KL}}}({{p}_{\pi }}(\tau )||p(\tau ))={{\mathbb{E}}_{\tau \sim{{p}_{\pi }}(\tau )}}[\log p(\tau )-\log {{p}_{\pi }}(\tau )] \\ 
& ={{\mathbb{E}}_{({{x}_{t}},{{u}_{t}})\sim{{p}_{\pi }}({{x}_{t}},{{u}_{t}})}}\left[ \log \eta ({{x}_{1}}) \right.\\
&\left. +\sum\limits_{t=1}^{T}{\left( \log p({{x}_{t+1}}|{{x}_{t}},{{u}_{t}})+r({{x}_{t}},{{u}_{t}}) \right)} -\log Z \right.\\ 
& \left. -\log \eta ({{x}_{1}})-\sum\limits_{t=1}^{T}{\left( \log p({{x}_{t+1}}|{{x}_{t}},{{u}_{t}})-\log \pi ({{u}_{t}}|{{x}_{t}}) \right)} \right] \\ 
& ={{\mathbb{E}}_{\pi }}\left[ \left( \sum\limits_{t=1}^{T}{r({{x}_{t}},{{u}_{t}})-\log \pi ({{u}_{t}}|{{x}_{t}})} \right)-\log Z \right] \\ 
& \propto \sum\limits_{t=1}^{T}{{{\mathbb{E}}_{({{x}_{t}},{{u}_{t}})\sim{{p}_{\pi }}({{x}_{t}},{{u}_{t}})}}}[r({{x}_{t}},{{u}_{t}})-\log \pi ({{u}_{t}}|{{x}_{t}})] \\ 
& =\sum\limits_{t=1}^{T}{{{\mathbb{E}}_{({{x}_{t}},{{u}_{t}})\sim{{\rho }_{\pi }}}}}[r({{x}_{t}},{{u}_{t}})+\mathcal{H}(\pi ({{u}_{t}}|{{x}_{t}}))].
\end{aligned}\label{eq:5}
\end{equation}

Therefore, the objective of MaxEnt RL is to maximize the entropy-regularized long-term (discounted) rewards as shown in Eq. \eqref{eq:2}. From another perspective, this is equivalent to minimizing the KL divergence between 1) the target (optimal) trajectory distribution $p(\tau )$ induced by $r({x_t},{u_t})$, and 2) the trajectory distribution ${p_\pi }(\tau )$ generated by the MaxEnt RL policy $\pi $ that is to be learned. The relationship presented in Eq. \eqref{eq:5} can be formally expressed as follows:
\begin{equation}
\begin{aligned}
&\arg \underset{\pi }{\mathop{\min }}\,{{D}_{\text{KL}}}({{p}_{\pi }}(\tau )||p(\tau ))\\
&=\arg \underset{\pi }{\mathop{\max }}\,\sum\limits_{t=1}^{T}{{{\mathbb{E}}_{({{x}_{t}},{{u}_{t}})\sim{{\rho }_{\pi }}}}}[r({{x}_{t}},{{u}_{t}})+\mathcal{H}(\pi ({{u}_{t}}|{{x}_{t}}))].
\end{aligned}
\end{equation}

\subsection{Maximum entropy inverse RL}\noindent
Inverse RL aims to infer the intent (reward function) of an expert by observing its behaviors, specifically through optimal expert demonstrations (trajectories). MaxEnt IRL is a classical IRL method within the above MaxEnt RL framework \cite{RN13}, which simultaneously learns the reward function and the policy based on the trajectories ${p_{{\pi _E}}}(\tau )$ generated by a particular expert policy $\pi _E$. Formally, the objective is to infer a reward function ${r_\theta }({x_t},{u_t})$ parametrized by $\theta $, using the optimal trajectory distribution ${p_\theta }(\tau )$, similar to Eq. \eqref{eq:4}:
\begin{equation}\begin{array}{c}
{p_\theta }(\tau ) \buildrel \Delta \over = p(\tau |\theta ) = \frac{1}{{{Z_\theta }}}\eta ({x_1})\prod\limits_{t = 1}^T {p({x_{t + 1}}|{x_t},{u_t})} \exp ({r_\theta }({x_t},{u_t}))\\
\propto \left[ {\eta ({x_1})\prod\limits_{t = 1}^T p ({x_{t + 1}}|{x_t},{u_t})} \right]\exp \left( {\sum\limits_{t = 1}^T {{r_\theta }} ({x_t},{u_t})} \right)
\end{array}
\end{equation}
where the partition function is defined as:
\begin{equation}
{Z_\theta } \buildrel \Delta \over = \int {\eta ({x_1})\prod\nolimits_t {p({x_{t + 1}}|{x_t},{u_t})\exp ({r_\theta }({x_t},{u_t}))} d\tau }
\end{equation}
Then the objective of MaxEnt IRL is to address the maximum likelihood estimation (MLE) problem as follows:
\begin{equation}
\begin{aligned}
& \arg \underset{\theta }{\mathop{\min }}\,{{D}_{\text{KL}}}({{p}_{{{\pi }_{E}}}}(\tau )||{{p}_{\theta }}(\tau ))=\arg \underset{\theta }{\mathop{\max }}\,{{\mathbb{E}}_{{{p}_{{{\pi }_{E}}}}(\tau )}}\left[ \log {{p}_{\theta }}(\tau ) \right] \\ 
& ={{\mathbb{E}}_{\tau \sim{{\pi }_{E}}}}\left[ \sum\limits_{t=1}^{T}{{{r}_{\theta }}}({{x}_{t}},{{u}_{t}}) \right]-\log {{Z}_{\theta }}  
\end{aligned}\label{eq:9}
\end{equation}

\subsection{Adversarial inverse RL (AIRL)}\noindent
However, the term ${Z_\theta }$ in the MLE problem in Eq. \eqref{eq:9} can become computationally intractable when the state-action spaces are large or even continuous. Additionally, MaxEnt IRL requires explicit knowledge of the environment’s dynamics, which is often infeasible. To address these issues, adversarial IRL (AIRL) \cite{RN11} is proposed to cast optimization of Eq. \eqref{eq:9} as a GAN problem. This approach also extends conventional methods that learn from entire trajectories into just learning over single state-action pairs. The discriminator ${D_\theta }$ in the AIRL algorithm is chosen as a particular form:
\begin{equation}
{D_\theta }(x,u) = \frac{{\exp \{ {r_\theta }(x,u)\} }}{{\exp \{ {r_\theta }(x,u)\}  + {\pi _\omega }(u|x)}}\label{eq:10}
\end{equation}
where ${r_\theta }(x,u)$ is the learned reward function and ${\pi _\omega }(u|x)$ is the corresponding policy induced by the reward ${r_\theta }$. In Eq. \eqref{eq:10}, ${\pi _\omega }(u|x)$ is precomputed as a filled-in value for ${D_\theta }$. The discriminator aims to distinguish between the samples from the expert demonstrations and those generated by the current policy ${\pi _\omega }$. In the AIRL algorithm, the policy ${\pi _\omega }$ is trained to maximize ${{\mathbb{E}}_{{{\rho }_{{{\pi }_{\omega }}}}}}[\log {{D}_{\theta }}(x,u)-\log (1-{{D}_{\theta }}(x,u))]$, which is equivalent to maximizing the objective of an MaxEnt RL policy as follows:
\begin{equation}
\begin{aligned}
&{{\mathbb{E}}_{{{\pi }_{\omega }}}}\left[ \sum\limits_{t=1}^{T}{\log }({{D}_{\theta }}({{x}_{t}},{{u}_{t}}))-\log (1-{{D}_{\theta }}({{x}_{t}},{{u}_{t}})) \right] \\
&={{\mathbb{E}}_{{{\pi }_{\omega }}}}\left[ \sum\limits_{t=1}^{T}{{{r}_{\theta }}}({{x}_{t}},{{u}_{t}})-\log {{\pi }_{\omega }}({{u}_{t}}|{{x}_{t}}) \right]
\end{aligned}
\end{equation}
Therefore, the objective of the discriminator in the GAN-inspired IRL framework aligns precisely with learning the reward function. Simultaneously, the policy (generator) is adjusted to make it increasingly difficult for the discriminator to distinguish between expert demonstrations and samples generated by the policy. It can be demonstrated that, when trained to optimality, the learned reward function ${r_\theta }(x,u)$ can recover the ground-truth reward up to a constant, provided that the true reward is solely a function of state \cite{RN11}.

\section{Problem statement}\label{Problem_statement}
\subsection{Multi-mode process control problem}\label{Problem}\noindent
The general state-space representation of a control system can be described as follows:
\begin{equation}
{x_{t + 1}} \sim p({x_{t + 1}}|{x_t},{u_t})\label{eq:12}
\end{equation}
\begin{equation}
{u_t} \sim p({u_t}|{x_t},\omega ) \buildrel \Delta \over = {\pi _\omega }({u_t}|{x_t})\label{eq:13}
\end{equation}
where $p({u_t}|{x_t},\omega )$ is the conditional distribution of actions explicitly denoted as a $\omega$-parameterized policy ${\pi _\omega }({u_t}|{x_t})$ to emphasize the role of the control policy.

Utilizing the system dynamics model and controller described above, the evolution trajectory of the MDP unfolds from the initial state as follows:
\begin{equation}
\begin{aligned}
p(\tau ) & = p({x_1},{u_t}, \ldots ,{x_T},{u_T}|\omega ) \\
& = \eta ({x_1})\prod\limits_{t = 1}^T {p({x_{t + 1}}|{x_t},{u_t}){\pi _\omega }({u_t}|{x_t})} .
\end{aligned}
\end{equation}

However, in process control scenarios, many controlled processes naturally exhibit multi-mode behaviors. These processes often operate under a variety of modes or working conditions, which may include different optimization and control setpoints, varying feed compositions (recipes), distinct system parameters, and even different equipment scales. Such variability leads to different dynamic models for each operating mode. In this paper, the term ``multi-mode processes'' is used broadly to describe processes characterized by multiple distinct models from their corresponding operating scenarios.

Formally, each mode within a multi-mode control system possesses unique characteristics, reflecting inherent differences in their dynamics $p({x_{t + 1}}|{x_t},{u_t})$. Nevertheless, these operating modes also share certain common features, which adhere to an underlying probability distribution $p( \cdot | \cdot )$. Unlike the general control systems described in Eqs. \eqref{eq:12}-\eqref{eq:13}, the optimal controllers for each mode in multi-mode processes are not necessarily consistent. This inconsistency necessitates the development of rational and suitable mathematical formulations to accurately describe the control problem for multi-mode processes.

\begin{figure}
	\centering
	\includegraphics[width=1.0\linewidth]{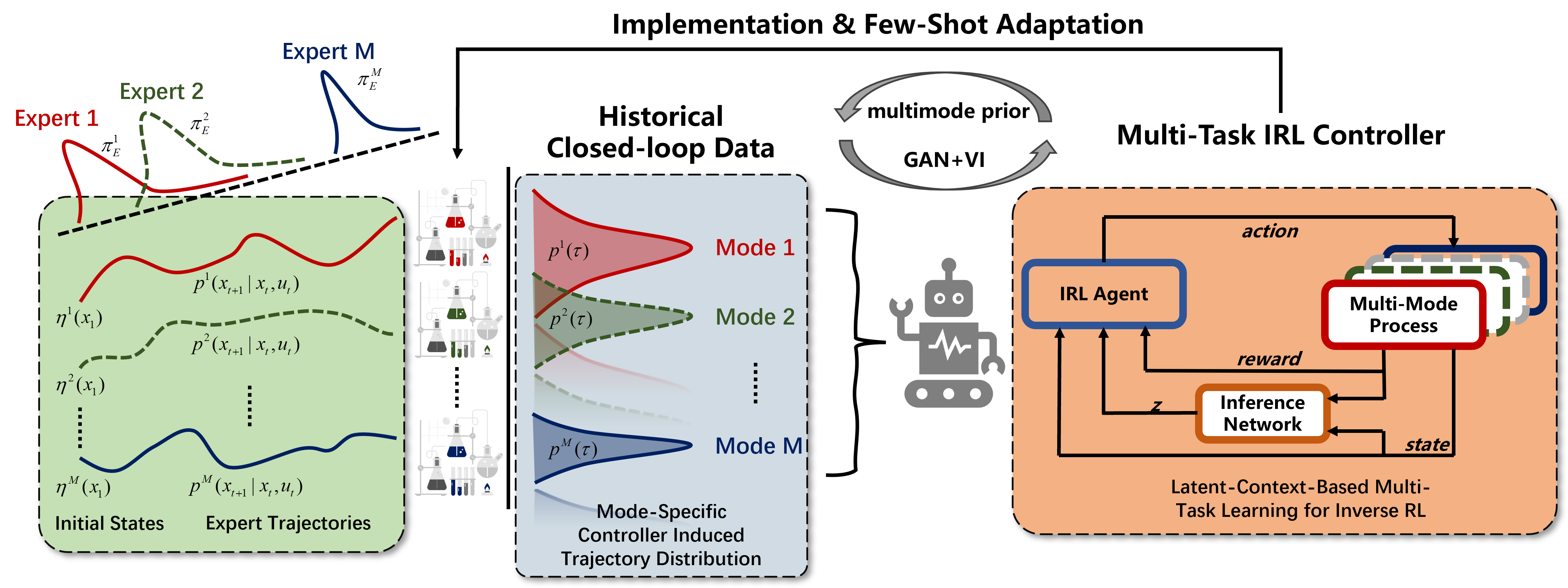}
	\caption{Multi-task inverse reinforcement learning framework for designing multi-mode process control systems.}
	\label{fig:fig-1}
\end{figure}

Given a process system with $M\in {{\mathbb{N}}^{+}}$ operating modes, each associated with a different optimal or near-optimal controller ${\pi _E} \buildrel \Delta \over = \{ \pi _E^1,\pi _E^2, \cdots ,\pi _E^M\} $, where the optimal controller for mode $m$ is denoted as $\pi _E^m$, the multi-mode controller ${\pi _E}$ will generate $M$ different but structurally similar trajectory distributions under the dynamics specific to each mode. Fig. \ref{fig:fig-1} presents the overall concept of the proposed framework. In multi-mode control systems, multiple expert controllers correspond to different operating modes, leading to distinct distributions of expert trajectories. While the trajectory of each mode captures its unique dynamic behavior, the overall closed-loop process control system exhibits inherent similarities across modes due to shared system architecture and underlying trajectory distributions.

Therefore, the multi-mode process control problem is to learn a universal controller ${\pi _\omega }({u_t}|{x_t})$ that can adapt to various control objectives across different operating modes. This paper aims to leverage historical industrial closed-loop big data and multi-task IRL to learn controller priors from mode-specific trajectory distributions, enabling efficient few-shot adaptation during implementation. In other words, ${\pi _\omega } \buildrel \Delta \over = \{ \pi _\omega ^1,\pi _\omega ^2, \cdots ,\pi _\omega ^M\} $ should be capable of providing optimal or near-optimal control policies for all ${p^m}({x_{t + 1}}|{x_t},{u_t})$, i.e., ${\pi _\omega } \buildrel\textstyle.\over= {\pi _E}$ should be approximately satisfied for each mode.

\subsection{Context-conditional multi-task controller learning problem}\noindent
With the vast amount of historical closed-loop operational data available in industrial production, leveraging this multi-mode data for offline controller design presents both significant challenges and exciting opportunities. Developing an initial controller from such data could greatly enhance the practicality and safety of DRL applications. This paper proposes a framework for designing a multi-mode process controller based on IRL. The objective is to develop a fully closed-loop-data-driven controller that can effectively adapt to different operating modes.

As analyzed in Section \ref{Problem}, assuming the controlled system operates under $M$ distinct modes, corresponding to $M$ distinct optimal or near-optimal controllers, the operation of this multi-mode process control system will generate a large volume of historical closed-loop data. Over time, this data will encapsulate $M$ different trajectory distributions, which are implicitly and structurally embedded within the database. The next step is to formally define the problem: \textit{how can a multi-task IRL approach be employed to effectively solve the multi-mode process control challenge?}

This work introduces the concept of contextual policy to reframe the multi-task RL (IRL) for multi-mode process control as the problem of solving a context-conditional MDP using a latent context model. In a multi-mode process control system with structurally similar dynamic models and controllers, the MDP is augmented by a latent context variable, capturing dependencies across operating modes. This approach enables the development of a universally structured controller that can adapt to all modes while accounting for the unique characteristics of each individual mode.

Let $\pi ({u_t}|{x_t},z)$ represent the controller for each operating mode. Unlike conventional RL and IRL, where the optimal control policy is defined solely based on the current state ${x_t}$, this framework incorporates a mode-specific representation. The latent context variable $z$ is introduced as an additional explicit dependency in the conditional policy $\pi ( \cdot | \cdot ,z)$ for each mode. By incorporating probabilistic latent variable modeling, the generative model for the expert trajectory $\tau _E^z \buildrel \Delta \over = {\{ {{{x}}_{1:T}},{{{u}}_{1:T}}\} _z}$ under the $z$-dependent operating mode can be expressed as:
\begin{equation}
\begin{aligned}
{x_1} \sim \eta ({x_1}),z \sim p(z),{u_t} \sim \pi ({u_t}|{x_t},z),{x_{t + 1}} \sim p({x_{t + 1}}|{x_t},{u_t})
\end{aligned}\label{eq:15}
\end{equation}
The joint distribution of the latent variable $z$ and the trajectory $\tau _E^z$ can be expressed as:
\begin{equation}
{p_{{\pi _E}}}(z,\tau ) = p(z){p_{{\pi _E}}}(\tau |z)
\end{equation}
The marginal distribution of the overall historical dataset, which consists of data from different modes, can be represented as:
\begin{equation}
{p_{{\pi _E}}}(\tau ) = \int_{\cal Z} {{p_{{\pi _E}}}(z,\tau )} dz = \int_{\cal Z} {p(z){p_{{\pi _E}}}(\tau |z)} dz\label{eq:17}
\end{equation}
where $p(z)$ represents the prior distribution of the latent context variable (i.e., the mode-indicating variable), $\eta ({x_1})$ and $p({x_{t + 1}}|{x_t},{u_t})$ respectively denote the probability distributions for the initial state and state transition, and ${p_{{\pi _E}}}(\tau |z)$ represents the context-conditional trajectory distribution.

The objective is to learn the control policies ${\pi _\omega } \buildrel \Delta \over = \{ \pi _\omega ^1,\pi _\omega ^2, \cdots ,\pi _\omega ^M\} $ for each mode, along with the corresponding optimality or rationality descriptions (which, in the context of DRL, represent the reward functions) based on the overall marginal distribution ${p_{{\pi _E}}}(\tau )$. However, the probabilistic latent variable formulation above abstracts a complex multi-mode process control problem. In real-world industrial processes, the necessary assumptions may not hold. Specifically, there is often limited or no knowledge of the prior distribution $p(z)$ representing the mode-indicating variable $z$ in Eq. \eqref{eq:15} and Eq. \eqref{eq:17}, and the conditional trajectory distribution ${p_{{\pi _E}}}(\tau |z)$ for different modes may be unknown. In other words, in practice, only the marginal distribution ${p_{{\pi _E}}}(\tau )$ of the entire set of expert trajectories is observable, which makes learning from multi-mode demonstrations particularly challenging.

\section{Methodology}\label{Methodology}
\subsection{From single-task to multi-task MDP}\noindent
To apply a multi-task IRL approach for learning controllers from historical closed-loop operating data of multi-mode processes, the conventional single-mode MDP definition must be extended. Specifically, the original MDP is modified and augmented by introducing a conditional term based on $z \in {\cal Z}$, where ${\cal Z}$ represents the value space of the latent context variable $z$. Consequently, each MDP component—except for the state transition probability determined by system dynamics—will now include an additional dependency on $z$. This enables the MDP to effectively capture the varying characteristics of different operating modes within the process.

Based on the latent context model, the context-conditional policy is now defined as $\pi :{\cal X} \times {\cal Z} \to {\cal P}({\cal U})$, and the corresponding reward function is modified to $r:\mathcal{X}\times \mathcal{U}\times \mathcal{Z}\to \mathbb{R}$. This generalized MDP formulation enables the controller design for each mode to be conditioned on the latent context $z$. In other words, with different predetermined values of $z$, the multi-task IRL agent can be trained in a mode-specific manner for various control tasks that share common structures or feature spaces, allowing for efficient adaptation across multiple operating modes.

Building upon the MaxEnt RL framework in Eq. \eqref{eq:2}, in multi-mode scenarios, the optimal context-conditional policy can be calculated as:
\begin{equation}
\begin{aligned}
\pi _{E}^{z}\leftarrow {{\pi }^{*}}=\arg \underset{\pi }{\mathop{\max }}\,{{\mathbb{E}}_{z \sim p(z),({{{x}}_{1:T}},{{{u}}_{1:T}})\sim{{p}_{\pi }}(\cdot |z)}}\\
\left[ \sum\limits_{t=1}^{T}{r({{x}_{t}},{{u}_{t}},z)-\log \pi ({{u}_{t}}|{{x}_{t}},z)} \right]
\end{aligned}
\end{equation}
where $r({x_t},{u_t},z)$ is the mode-specific reward function, and $ - \log \pi ({u_t}|{x_t},z)$ is the entropy-regularization term for the contextual policies. Analogous to Eq. \eqref{eq:3}, the context-conditional distribution for the $z$-th expert trajectory ${p_{{\pi _E}}}(\tau |z)$ in Eq. \eqref{eq:17} can be formulated as follows:
\begin{equation}
\begin{aligned}
\tau _E^z & \sim {p_{{\pi _E}}}(\tau |z) = {p_{{\pi _E}}}({{{x}}_{1:T}},{{{u}}_{1:T}}|z) \\
& = \eta ({x_1})\prod\limits_{t = 1}^T {p({x_{t + 1}}|{x_t},{u_t}){\pi _E}} ({u_t}|{x_t},z)
\end{aligned}\label{eq:19}
\end{equation}

In summary, the multi-mode process control problem in this paper is reframed as a context-conditioned multi-task IRL training approach that effectively learns from historical multi-mode closed-loop big data. This approach utilizes a set of multi-task demonstrations ${\tau _E}$ \textit{i.i.d.} sampled from the marginal distribution ${p_{{\pi _E}}}(\tau )$ defined by Eq. \eqref{eq:17} and Eq. \eqref{eq:19}, i.e.,
\begin{equation}
\begin{array}{c}
{\tau _E} \sim {p_{{\pi _E}}}(\tau ) = \int_{\cal Z} {p(z){p_{{\pi _E}}}(\tau |z)} dz\\
= \int_{\cal Z} {p(z)\eta ({x_1})\prod\limits_{t = 1}^T {p({x_{t + 1}}|{x_t},{u_t}){\pi _E}({u_t}|{x_t},z)} } dz
\end{array}
\end{equation}
The context-conditioned multi-task IRL approach aims to uncover historical multi-mode patterns and subsequently learn a multi-mode controller prior in a purely data-driven manner. This methodology enables the identification of distinct operating modes and their corresponding control strategies, supporting the development of a robust controller capable of adapting to varying conditions using only historical data.

\subsection{Latent context inference model for multi-task learning}\noindent
Since both the reward function and the control policy are conditioned on $z$ in the generalized MDP, estimating this latent variable is essential for multi-task IRL-based controller learning. To achieve this, a probabilistic inference model $q(z|\tau )$ should be introduced to approximate the true posterior distribution $p(z|\tau  = \tau _E^z)$, which is generally inaccessible. By denoting the inference model as a variational approximation for calculating $z$, the context-conditional reward function ${r_{{\rm{IRL}}}}(x,u,z)$ (which needs to be learned) can be determined using the inferred $z$.

Specifically, when given a set of demonstrations sampled from the mode prior and the conditional distribution $z \sim p(z),\tau _E^z \sim {p_{{\pi _E}}}(\tau |z)$, the inference model can be employed to estimate the latent context variable $\hat z \sim q(z|\tau _E^z)$. Once the mode-indicating variable $\hat z$ is inferred, it can be substituted into the learned reward ${r_{{\rm{IRL}}}}(x,u,\hat z)$. The DRL agent, guided by this context-conditional reward, should then generate policies that closely resemble those driven by the true underlying reward $r(x,u,z)$. Successfully training the multi-task IRL agent with latent dependencies equips the IRL-based process controller to effectively manage scenarios characterized by multi-mode behaviors.

\subsection{Multi-task IRL using context-conditional probabilistic inference}\noindent
The next step involves analyzing how to address the multi-task IRL problem to facilitate the learning of a multi-mode process controller. Drawing from the MLE approach for MaxEnt IRL outlined in Section \ref{Preliminaries} (Eq. \eqref{eq:9}), and considering the previously defined multi-task IRL problem based on latent context model, the context-conditional trajectory distribution, parameterized by the reward parameter $\theta$, can be derived as follows:
\begin{equation}
\begin{array}{c}
\tau _\theta ^z \sim {p_\theta }(\tau |z) = {p_\theta }({{{x}}_{1:T}},{{{u}}_{1:T}}|z)\\
= \frac{1}{{{Z_\theta }}}\left[ {\eta ({x_1})p({x_{t + 1}}|{x_t},{u_t})} \right]\exp \left( {\sum\limits_{t = 1}^T {{r_\theta }({x_t},{u_t},z)} } \right)
\end{array}\label{eq:21}
\end{equation}
where the conditional input $z$ is inferred using an inference model ${q_\psi }(z|\tau )$ parametrized by $\psi $, ${r_\theta }$ is the reward function in the multi-task IRL, while ${Z_\theta }$ is the partition function. To develop a multi-mode process controller, this framework ensures that the context-conditional trajectory distribution accurately captures the dependencies imposed by the latent context.

Therefore, the primary goal of multi-task IRL is to solve an MLE problem:
\begin{equation}
\begin{aligned}
& \arg \underset{\theta }{\mathop{\min }}\,{{\mathbb{E}}_{p(z)}}\left[ {{D}_{\text{KL}}}({{p}_{{{\pi }_{E}}}}(\tau |z)||{{p}_{\theta }}(\tau |z)) \right] \\ 
& =\arg \underset{\theta }{\mathop{\max }}\,{{\mathbb{E}}_{p(z),{{p}_{{{\pi }_{E}}}}(\tau |z)}}\left[ \log {{p}_{\theta }}(\tau |z) \right] \\
& =\arg \underset{\theta }{\mathop{\max }}\,{{\mathbb{E}}_{z \sim p(z),\tau \sim\pi _{E}^{z}}}\left[ \sum\limits_{t=1}^{T}{{{r}_{\theta }}}({{x}_{t}},{{u}_{t}},z) \right]-\log {{Z}_{\theta }} \\ 
\end{aligned}\label{eq:22}
\end{equation}
To achieve this, various IRL algorithms, such as AIRL, can be employed to minimize the KL divergence between the optimal or near-optimal expert trajectory distribution and the $\theta  $-induced trajectory distribution. The underlying objective of IRL agent is to match the trajectory distribution of $\tau  = \{ {{{x}}_{1:T}},{{{u}}_{1:T}}\} $. However, the challenge arises because every trainable term in Eq. \eqref{eq:22} is conditioned on the latent context $z$, which is not directly related to the IRL optimization process. This limitation complicates the learning process, as the IRL agent struggles to effectively optimize distinct modes without a well-defined association between trajectory distributions and latent contexts. In other words, an explicit correlation between the $\theta $-induced trajectory $\tau $ and the latent variable $z$ must be enforced to distinguish the conditional reward function and the corresponding policy within each mode.

Building on the principles of InfoGAN \cite{RN9}, mutual information (MI) between $z$ and $\tau $ can be applied as a constraint or correlation measure to enhance the dependency between the latent context and the resulting trajectory. Within the framework of multi-task IRL, a higher MI value indicates a stronger correlation, thereby improving the interpretability of $z$ in relation to the trajectory ${\tau ^z}$. The MI under joint distribution ${p_\theta }(z,\tau )$ is calculated as follows:
\begin{equation}
\begin{aligned}
& {{I}_{{{p}_{\theta }}}}(z;\tau )=\mathcal{H}(z)-\mathcal{H}(z|\tau ) \\ 
& =\mathcal{H}(z)+{{\int_{z}{\int_{\tau }{p}_{\theta }}}}(z,\tau )\log {{p}_{\theta }}(z,\tau )dzd\tau  \\ 
& =\mathcal{H}(z)+\int_{z}{\int_{\tau }{p}}(z){{p}_{\theta }}(\tau |z)\log {{p}_{\theta }}(z|\tau )dzd\tau  \\ 
& =\mathcal{H}(z)+{{\mathbb{E}}_{z \sim p(z),\tau \sim{{p}_{\theta }}(\tau |z)}}[\log {{p}_{\theta }}(z|\tau )] \\ 
& ={{\mathbb{E}}_{z \sim p(z),\tau \sim{{p}_{\theta }}(\tau |z)}}[\log {{p}_{\theta }}(z|\tau )-\log p(z)]  
\end{aligned}\label{eq:23}
\end{equation}
Since ${p_\theta }(z|\tau )$ is the posterior distribution that is unknown, the probabilistic inference model ${q_\psi }(z|\tau )$ can serve as a variational approximation of this posterior, and then Eq. \eqref{eq:23} can be reformulated as:
\begin{equation}
\begin{aligned}
& {{I}_{{{p}_{\theta }}}}(z;\tau )=\mathcal{H}(z)+{{\mathbb{E}}_{z \sim p(z),\tau \sim{{p}_{\theta }}(\tau |z)}}[\log {{p}_{\theta }}(z|\tau )] \\ 
& =\mathcal{H}(z)+{{\mathbb{E}}_{z \sim p(z),\tau \sim{{p}_{\theta }}(\tau |z)}}[\underbrace{{{D}_{\text{KL}}}({{p}_{\theta }}||{{q}_{\psi }})}_{\ge 0}+\log {{q}_{\psi }}(z|\tau )] \\ 
& \ge \mathcal{H}(z)+{{\mathbb{E}}_{z \sim p(z),\tau \sim{{p}_{\theta }}(\tau |z)}}[\log {{q}_{\psi }}(z|\tau )] \\ 
& ={{\mathbb{E}}_{z \sim p(z),\tau \sim{{p}_{\theta }}(\tau |z)}}[\log {{q}_{\psi }}(z|\tau )-\log p(z)] \\ 
& ={{L}_{I}}({{p}_{\theta }},{{q}_{\psi }})  
\end{aligned}\label{eq:24}
\end{equation}
where ${L_I}({p_\theta },{q_\psi })$ is the variational lower bound of the MI.

The following analysis will outline the solution of the multi-task IRL problem. In this context, Eq. \eqref{eq:22} represents the primary objective based on the MaxEnt principle, while Eq. \eqref{eq:24} introduces an additional regularization term for the latent context variable. Accordingly, the overall optimization objective can be formulated as:
\begin{equation}
\begin{aligned}
& \underset{\theta ,\psi }{\mathop{\min }}\,{{\mathbb{E}}_{p(z)}}[{{D}_{\text{KL}}}({{p}_{{{\pi }_{E}}}}(\tau |z)||{{p}_{\theta }}(\tau |z))] \\ 
& -\alpha \cdot {{I}_{{{p}_{\theta }}}}(z;\tau )+\beta \cdot {{\mathbb{E}}_{{{p}_{\theta }}(\tau )}}[{{D}_{\text{KL}}}({{p}_{\theta }}(z|\tau )||{{q}_{\psi }}(z|\tau ))] \\ 
\end{aligned}\label{eq:25}
\end{equation}
The first term aims to align the conditional distributions between the closed-loop expert trajectories and the IRL trajectories generated by the $\theta $-parameterized reward function and the corresponding RL policy. This alignment represents the primary objective of the context-conditional multi-task IRL problem within the MaxEnt reward learning framework. The second term seeks to maximize the MI between the context and the corresponding trajectory, ensuring that the information embedded in the latent context $z$ is retained throughout the training process \cite{RN9}. Finally, the third term addresses the alignment of the variational inference approximation ${q_\psi }(z|\tau )$ with the true posterior ${p_\theta }(z|\tau )$ of the latent context, which is required to train the inference model ${q_\psi }$.

For simplicity and without loss of generality, the tunable hyperparameters can be treated as constants $\alpha  = \beta  = 1$. Consequently, Eq. \eqref{eq:25} can be expressed as:
\begin{equation}
\begin{aligned}
& \underset{\theta ,\psi }{\mathop{\min }}\,{{\mathbb{E}}_{p(z)}}\left[ {{D}_{\text{KL}}}({{p}_{{{\pi }_{E}}}}(\tau |z)||{{p}_{\theta }}(\tau |z)) \right] \\
& +{{\mathbb{E}}_{z \sim p(z),\tau \sim{{p}_{\theta }}(\tau |z)}}\left[ \log \frac{p(z)}{{{p}_{\theta }}(z|\tau )}+\log \frac{{{p}_{\theta }}(z|\tau )}{{{q}_{\psi }}(z|\tau )} \right] \\ 
& \equiv \underset{\theta ,\psi }{\mathop{\max }}\,-{{\mathbb{E}}_{p(z)}}\left[ {{D}_{\text{KL}}}({{p}_{{{\pi }_{E}}}}(\tau |z)||{{p}_{\theta }}(\tau |z)) \right] \\
& +\underbrace{{{\mathbb{E}}_{z \sim p(z),\tau \sim{{p}_{\theta }}(\tau |z)}}\left[ \log {{q}_{\psi }}(z|\tau )-\log p(z) \right]}_{{{L}_{I}}({{p}_{\theta }},{{q}_{\psi }})} \\ 
& =\underset{\theta ,\psi }{\mathop{\max }}\,-{{\mathbb{E}}_{p(z)}}\left[ {{D}_{\text{KL}}}({{p}_{{{\pi }_{E}}}}(\tau |z)||{{p}_{\theta }}(\tau |z))-\underbrace{\log p(z)}_{\text{regardless of }\theta ,\psi } \right]\\
& +{{\mathbb{E}}_{z \sim p(z),\tau \sim{{p}_{\theta }}(\tau |z)}}\log {{q}_{\psi }}(z|\tau ) \\ 
& =\underset{\theta ,\psi }{\mathop{\max }}\,-{{\mathbb{E}}_{p(z)}}\left[ {{D}_{\text{KL}}}({{p}_{{{\pi }_{E}}}}(\tau |z)||{{p}_{\theta }}(\tau |z)) \right]\\
& +{{\mathbb{E}}_{z \sim p(z),\tau \sim{{p}_{\theta }}(\tau |z)}}\log {{q}_{\psi }}(z|\tau ) \\ 
& =\underset{\theta ,\psi }{\mathop{\max }}\,-{{\mathbb{E}}_{p(z)}}\left[ {{D}_{\text{KL}}}({{p}_{{{\pi }_{E}}}}(\tau |z)||{{p}_{\theta }}(\tau |z)) \right]+{{\mathcal{L}}_{\text{info}}}(\theta ,\psi ). \\ 
\end{aligned}\label{eq:26}
\end{equation}

\subsection{Practical implementation for solving multi-task IRL}\noindent
At this stage, the optimization objective in Eq. \eqref{eq:26} remains intractable, as it is not feasible to approximate the prior distribution $p(z)$ and the conditional trajectory distribution ${p_\theta }(\tau |z)$ by directly sampling from the marginal distribution ${p_{{\pi _E}}}(\tau )$ (i.e., expert trajectory distribution), which encompasses multi-mode process characteristics.

Fortunately, with the latent context inference model ${q_\psi }(z|\tau )$, an estimate over the sampled expert trajectory can be used to approximate the prior distribution $p(z)$ as follows:
\begin{equation}
{\tau _E} \sim {p_{{\pi _E}}}(\tau ),z \sim p(z) \buildrel\textstyle.\over= {q_\psi }(z|{\tau _E})
\end{equation}
And the conditional trajectory distribution ${p_\theta }(\tau |z)$ can be sampled from trajectories generated during the training process of the forward DRL agent within the inner loop of the multi-task IRL algorithm. This is feasible because, if the forward DRL policy ${\pi _\omega }$ is trained to optimality, the resulting trajectory distribution ${p_{\pi _\omega ^*}}(\tau |z)$ induced by the optimal policy $\pi _\omega ^ * $ will match the conditional trajectory distribution ${p_\theta }(\tau |z)$ \cite{RN15}.

With the approximately sampled $p(z)$ and ${p_\theta }(\tau |z)$, the second term in Eq. \eqref{eq:26} can be optimized with respect to $\theta $ and $\psi $. For the first KL divergence minimization term, any adversarial IRL algorithms such as AIRL can be applied. The only adjustment needed is to augment the RL state with an additional input dependency—the latent context $z$. This modification allows the IRL policy to be conditioned to  ${{\pi }_{\omega }}({{u}_{t}}|{{x}_{t}},z)$, where $\left\langle {x,z} \right\rangle $ serves as the augmented MDP state in the practical implementation of the algorithm.

Based upon the above analysis, the overall objective of the context-conditional multi-task IRL algorithm can be expressed as follows:
\begin{equation}
\begin{aligned}
& \underset{\omega }{\mathop{\min }}\,\underset{\theta ,\psi }{\mathop{\max }}\,{{\mathbb{E}}_{{{p}_{{{\pi }_{E}}}}(\tau ),{{q}_{\psi }}(z|{{\tau }_{E}}),{{\rho }_{{{\pi }_{\omega }}}}(x,u|z)}}\log (1-{{D}_{\theta }}(x,u,z)) \\
& +~{{\mathbb{E}}_{{{\tau }_{E}}\sim{{p}_{{{\pi }_{E}}}}(\tau ),z\sim{{q}_{\psi }}(z|{{\tau }_{E}})}}\log ({{D}_{\theta }}(x,u,m))+{{\mathcal{L}}_{\text{info}}}(\theta ,\psi ) \\
\end{aligned}
\end{equation}
where
\begin{equation}
{D_\theta }(x,u,z) = \frac{{\exp \{ {r_\theta }(x,u,z)\} }}{{\exp \{ {r_\theta }(x,u,z)\}  + {\pi _\omega }(u|x,z)}}
\end{equation}

\begin{figure}
	\centering
	\includegraphics[width=0.9\linewidth]{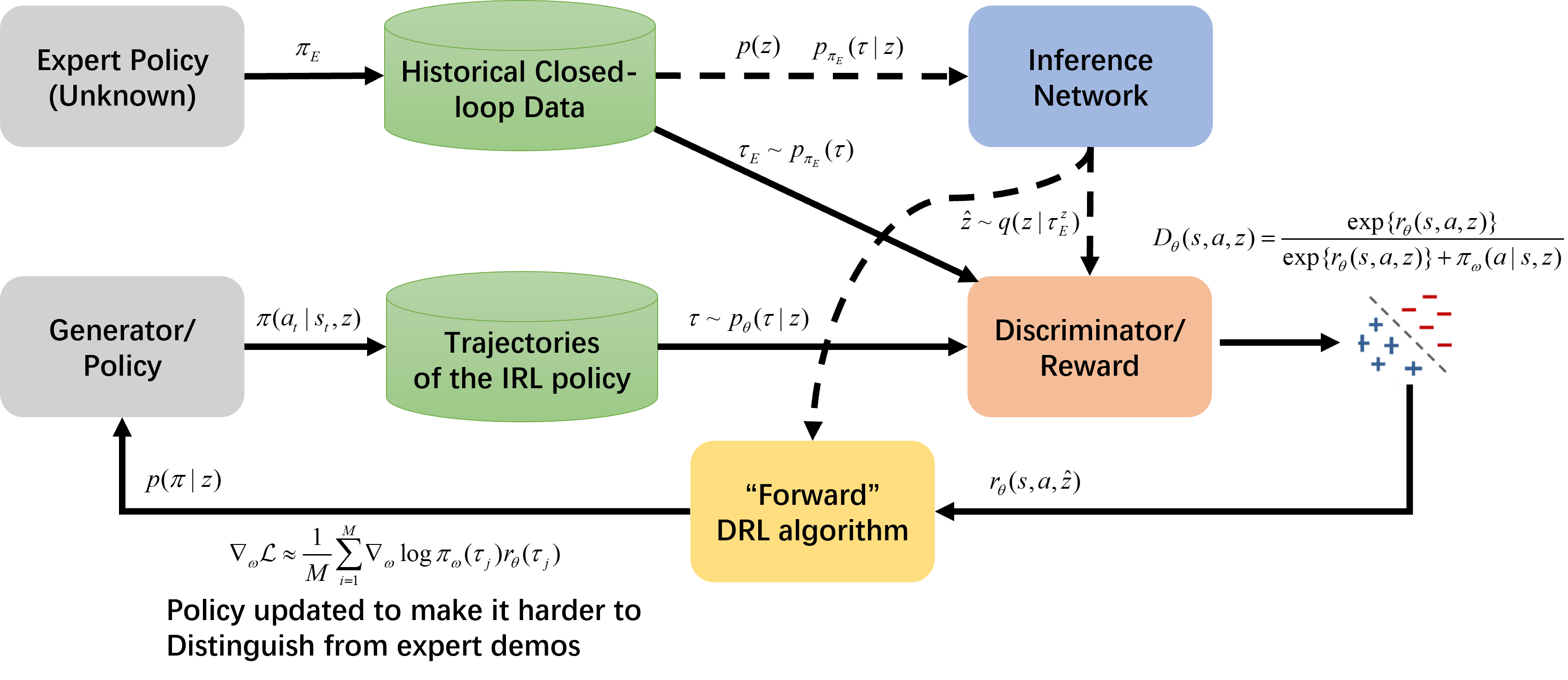}
	\caption{Flowchart of the proposed multi-task inverse reinforcement learning scheme.}
	\label{fig:fig-2}
\end{figure}

Therefore, the multi-task learning procedure for addressing the multi-mode process control problem can be summarized in Fig. \ref{fig:fig-2} and Algorithm~\ref{algorithm}. This approach involves three neural networks: the Generator (policy), the Discriminator (reward) and the Inference Network (mode indicator). Unlike conventional IRL, the proposed approach incorporates an additional input dependency, \textit{i.e.}, the latent context $z$, resulting in an augmented MDP state $\left\langle {x,z} \right\rangle $ that is used to train both the multi-mode policy and the reward. The process begins with the Inference Network estimating the mode-specific latent context of sampled trajectories. This inferred context is then fed into both the policy and the reward function. The GAN-inspired multi-task IRL framework utilizes a Generator to create virtual data corresponding to the current multi-mode policy, while the Discriminator defines the reward function by distinguishing expert data from the generated virtual data. During training, the outermost loop follows a GAN structure, where the Generator, acting as a DRL agent, updates the weights of the context-conditional policy based on the latest reward. The internal loop consists of any standard DRL algorithm, coupled with a parallel mode inference module, ensuring efficient learning across multiple modes. From the process control perspective, once the training on multi-mode historical closed-loop data is completed, the multi-task IRL-based controller can serve as an initialized multi-mode controller, enabling adaptability across those unseen operating modes in transfer learning settings.

\begin{algorithm}[h]
	\caption{Multi-task IRL training procedure}
	\footnotesize
	\label{algorithm}
	\begin{algorithmic}
		\STATE {\bfseries Input:} Expert trajectories $\mathcal{D}_E = \{\tau_E^j\}$; Initial parameters of $f_\theta, \pi_\omega, q_\psi$.
		\REPEAT
		\STATE Sample two batches of unlabeled demonstrations: $\tau_E, \tau'_E \sim \mathcal{D}_E$
		\STATE Infer a batch of latent context variables from the sampled demonstrations: $z \sim q_\psi(z|\tau_E)$
		\STATE Sample trajectories $\mathcal D$ from $\pi_\omega(\tau|z)$, with the latent context variable fixed during each rollout and included in $\mathcal D$.
		\STATE Update $\psi$ to increase $\mathcal L_\text{info}(\theta, \psi)$ with gradients in Eq. \eqref{eq:26}, with samples from $\mathcal D$.
		\STATE Update $\theta$ to increase $\mathcal L_\text{info}(\theta, \psi)$ with gradients in Eq. \eqref{eq:26}, with samples from $\mathcal D$.
		\STATE Update $\theta$ to decrease the binary classification loss:
		\STATE \begin{center}
			$\mathbb{E}_{(x,u,z) \sim \mathcal D}[\nabla_\theta \log D_\theta(x,u,z)] + \mathbb{E}_{\tau'_E \sim \mathcal D_E, z \sim q_\psi(z|\tau'_E)}[\nabla_\theta \log(1 - D_\theta(x,u,z))]$
		\end{center}
		\STATE Update $\omega$ with ``forward'' RL to increase the following objective:
		\STATE \begin{center} $\mathbb{E}_{(x,u,z) \sim \mathcal D} [\log D_\theta(x,u,z)]$\end{center}
		\UNTIL{Convergence}
		\STATE {\bfseries Output:} Learned inference model $q_\psi(z|\tau)$, reward function $f_\theta(x,u,z)$ and policy $\pi_\omega(u|x, z)$.
	\end{algorithmic}
\end{algorithm}

\section{Results \& discussion}\label{Experiment}
In this section, the proposed multi-mode controller learning method is applied to two distinct cases. The first case involves a fed-batch bioreactor, where the modes are characterized by unique system dynamics. The second case examines a continuous reactor, with modes defined by varying temperature setpoints.

\subsection{Case 1: A fed-batch bioreactor (batch profile optimization)}\noindent
To validate the feasibility of the multi-task IRL solution in recovering the reward function and addressing imitation learning control policies, this case study uses the same photo-production system as the numerical example presented in \cite{RN17}. The process involves a fed-batch bioreactor that necessitates solving a batch-to-batch optimization problem. Two distinct operating modes are defined, each corresponding to a unique set of internal system parameters. To implement this, the dynamic model is modified to incorporate a mode variable:
\begin{equation}\begin{array}{l}
\frac{{d{y_1}}}{{dt}} =  - \left( {{u_1} + 0.5u_1^2} \right){y_1} + {u_2}\\
\frac{{d{y_2}}}{{dt}} = {u_1}{y_1} - k \cdot {u_2}{y_1}
\end{array}\end{equation}
where ${u_1},{u_2}$ are the manipulated variables (i.e., light and an inflow rate) and ${y_1},{y_2}$ are the outlet concentrations of the reactant and product, respectively. The mode variable $k$ is:
\begin{equation}k = \left\{ \begin{array}{l}
0.5, \quad {\rm{Mode \ 1}}\\
0.7, \quad {\rm{Mode \ 2}}
\end{array} \right.
\end{equation}
The batch operation time course is normalized to 1, with control actions constrained within the interval [0, 5].

The objective is to design a control policy that adjusts the system inputs ${u_1},{u_2}$ to maximize the product concentration ${y_2}$ at the end of the batch operation. Positive reward feedback is only provided at the end of the batch operation, with rewards set to penalize excessive action changes at all other intervals. The reward function is defined as follows:
\begin{equation}
\begin{array}{l}
{r_t} = -0.01\times {{\left\| \mathbf{u}(t+1)-\mathbf{u}(t) \right\|}_{1}}, \quad t = 0,1, \cdots ,T - 1\\
{r_T} = {y_2}(T).
\end{array}\label{eq:31}
\end{equation}
The aim is to train an IRL agent to autonomously discover the optimal control policy from expert trajectories, ensuring that the recovered reward function aligns with the true reward function as described.

It is important to note that the experimental setup is highly challenging for both RL and IRL due to the sparsity of rewards, since positive feedback is granted only at the end of the episode. In this context, if the multi-task IRL agent fails to accurately infer the reward structures across different modes, it risks generating ineffective control policies. This may, in turn, lead to inaccurate reward estimates, perpetuating a feedback loop that hinders learning. Therefore, this setup serves as an ideal testbed for rigorously assessing the effectiveness of the multi-task IRL approach.

\begin{figure}
	\centering
	\includegraphics[width=1.0\linewidth]{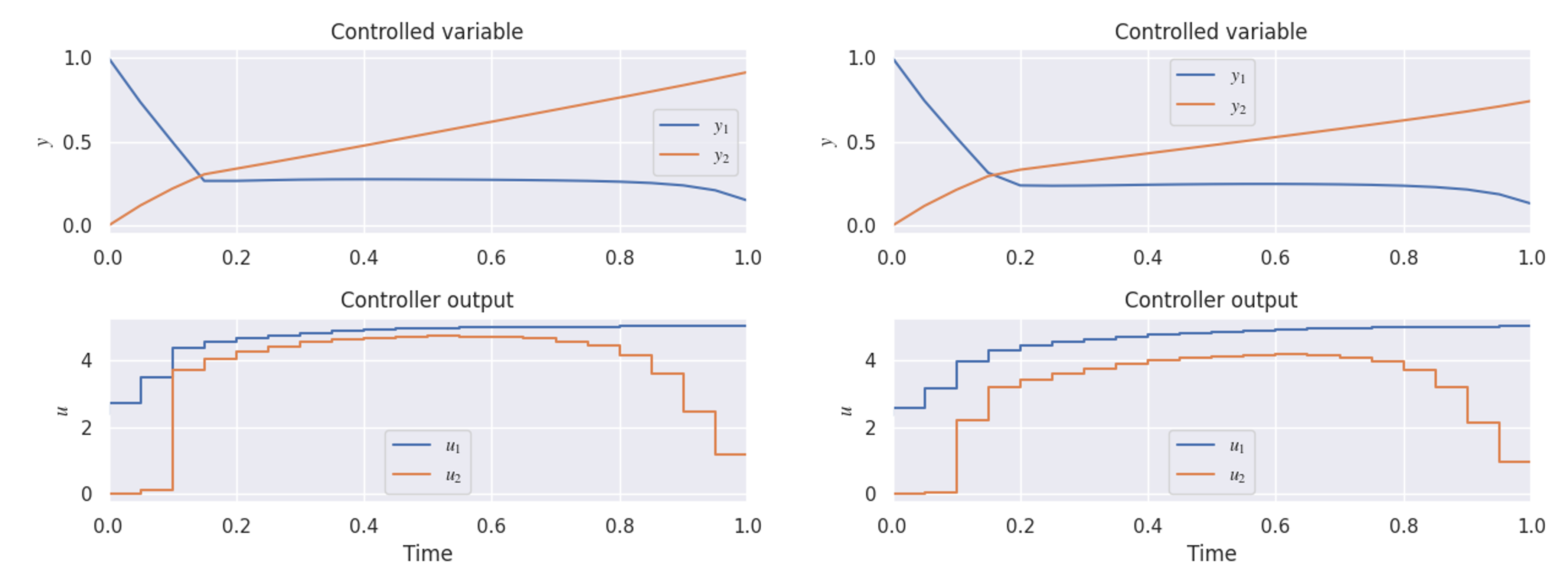}
	\caption{Typical batch optimization profiles of the TRPO expert demonstrations (left: Mode 1 $k = 0.5$; right: Mode 2 $k = 0.7$).}
	\label{fig:5.1}
\end{figure}

\begin{figure}
	\centering
	\includegraphics[width=1.0\linewidth]{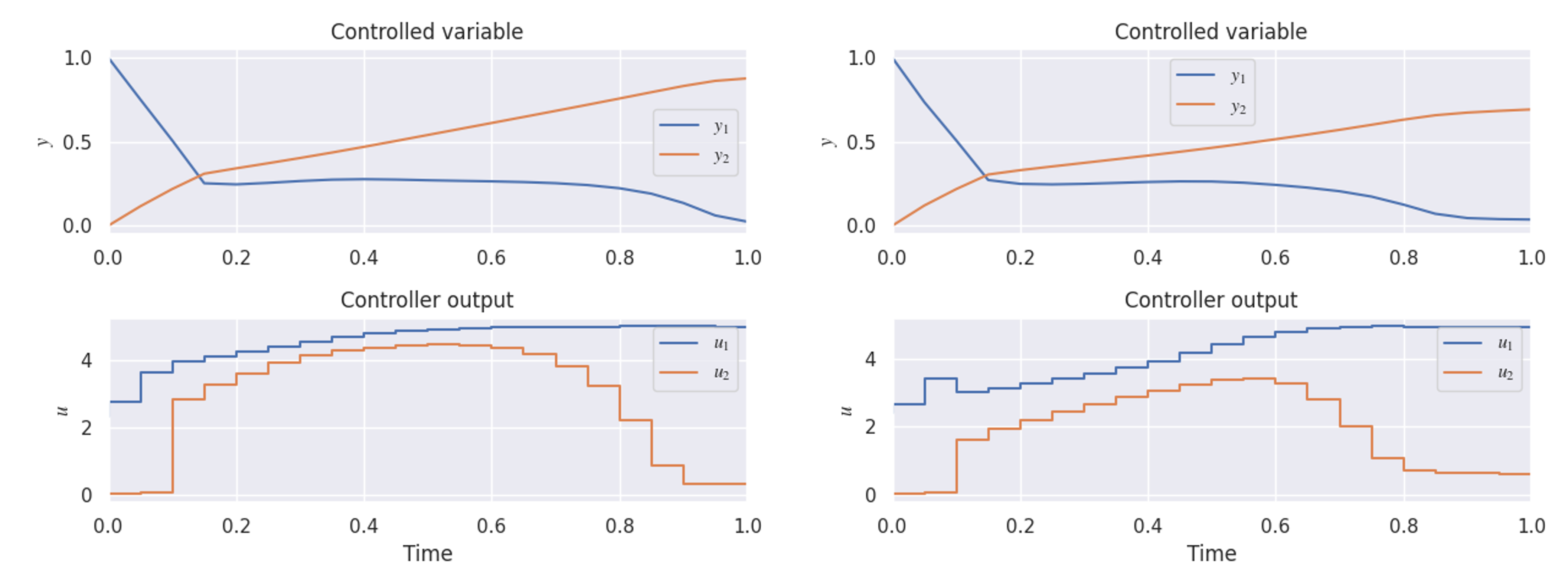}
	\caption{Batch optimization profiles of the successfully trained multi-task IRL agent based on the TRPO expert demonstrations (left: Mode 1 $k = 0.5$; right: Mode 2 $k = 0.7$).}
	\label{fig:5.2}
\end{figure}

First, a DRL-based expert policy is developed using the Trust Region Policy Optimization (TRPO) algorithm to maximize the reward function outlined in Eq. \eqref{eq:31}. In the RL setup, the state is represented as ${\cal X} \buildrel \Delta \over = {[{y_1},{y_2}]^T}$, and the action as ${\cal U} \buildrel \Delta \over = {[{u_1},{u_2}]^T}$. During training, each episode randomly selects an operating mode as the environment. The trained TRPO agent then generates expert demonstrations across both modes. Specifically, 2,112 expert trajectories are collected, with each comprising 20 samples. These trajectories are shuffled to simulate the mixed data sources typical in industrial production involving multiple devices. The typical expert trajectories for the two modes, $k = 0.5$ and $k = 0.7$, are illustrated in Fig. \ref{fig:5.1}, while the trajectories generated by the successfully trained multi-task IRL agent are displayed in Fig. \ref{fig:5.2}.


The results indicate that the multi-task IRL agent effectively recovers both the control policy and reward function from historical multi-mode data. At the endpoint of the batch trajectories, the product concentrations for each mode differ, reflecting the agent’s ability to learn the batch-to-batch optimization patterns unique to each operating mode. It is worth noting that, because the multi-task IRL method learns the reward function from historical data, it assumes expert demonstrations as optimal. Consequently, the recovered reward values are slightly lower than those of the expert, which is expected. This approach, centered on purely offline training of the multi-mode controller, is designed to fully leverage the prior knowledge embedded in expert behaviors. Such learned controller prior(s) can provide a strong foundation for subsequent transfer learning applications.

\subsection{Case 2: A benchmark CSTR process (continuous control)}
\subsubsection{System description and problem formulation}
\begin{figure}
	\centering
	\includegraphics[width=0.7\linewidth]{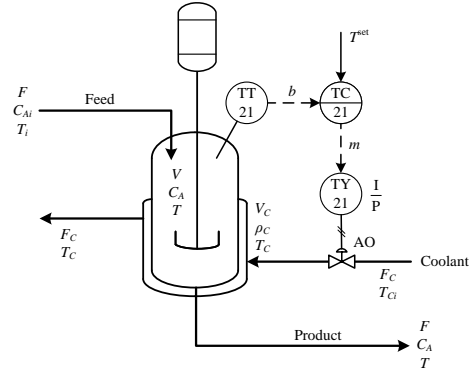}
	\caption{Sketch of the CSTR control system.}
	\label{fig:5.3}
\end{figure}

\noindent
To demonstrate the effectiveness of the proposed method in continuous control scenarios, a continuous stirred tank reactor (CSTR) process is selected as the test case, depicted in Fig. \ref{fig:5.3}. This reactor operates as a jacketed, non-adiabatic tank that facilitates a single irreversible and exothermic first-order reaction. The primary control objective is to maintain the reaction temperature $T$ near the target setpoint $T^{{\rm{set}}}$ by adjusting the valve opening $m$, which modulates the coolant feed flow rate. Further details on the first-principles model and system parameters are provided in \cite{RN80}.

Unlike the batch-to-batch optimization case, continuous processes require dynamic control solutions. Designing control strategies for continuous systems is considerably more complex than for batch profile optimization. In this scenario, the IRL agent faces increased difficulty due to the slow dynamic characteristics inherent to continuous process control. This necessitates a larger number of samples to effectively learn the reward function and establish a robust control policy, while also increasing instability in the training process.

In the experiment, two distinct modes are introduced to represent the multi-mode control scenario: Mode 1: Setpoint $88 \rightarrow 90$ \celsius, and Mode 2: Setpoint $88 \rightarrow 86$ \celsius. Expert trajectories from these modes are combined and randomly shuffled to mimic the multi-mode nature of industrial big data. The IRL agent is tasked with learning the characteristics of the expert controller from a multi-mode dataset with varied distributions, aiming to recover a reward function that can explain expert behavior(s). Two types of expert sources are used to validate the effectiveness of the multi-task IRL approach. The first type consists of expert demonstrations directly generated by a DRL agent. The second type uses multi-mode industrial closed-loop data as the basis for expert trajectories.

In this case, the RL state is represented as ${\cal X} \buildrel \Delta \over = {[{C_A},T,{T_C},b,T^{\rm{set}} - T]^T}$, and the action as ${\cal U} \buildrel \Delta \over = [m]$. For the multi-mode CSTR control problem, the temperature setpoint serves as the mode-indicating variable, which is unknown to the agent. Consequently, the inference model is critical for distinguishing between different operating modes within the encapsulated mode-specific information.

\subsubsection{DRL agent as optimal policy for generating expert demonstrations}
\begin{figure}
	\centering
	\includegraphics[width=1.0\linewidth]{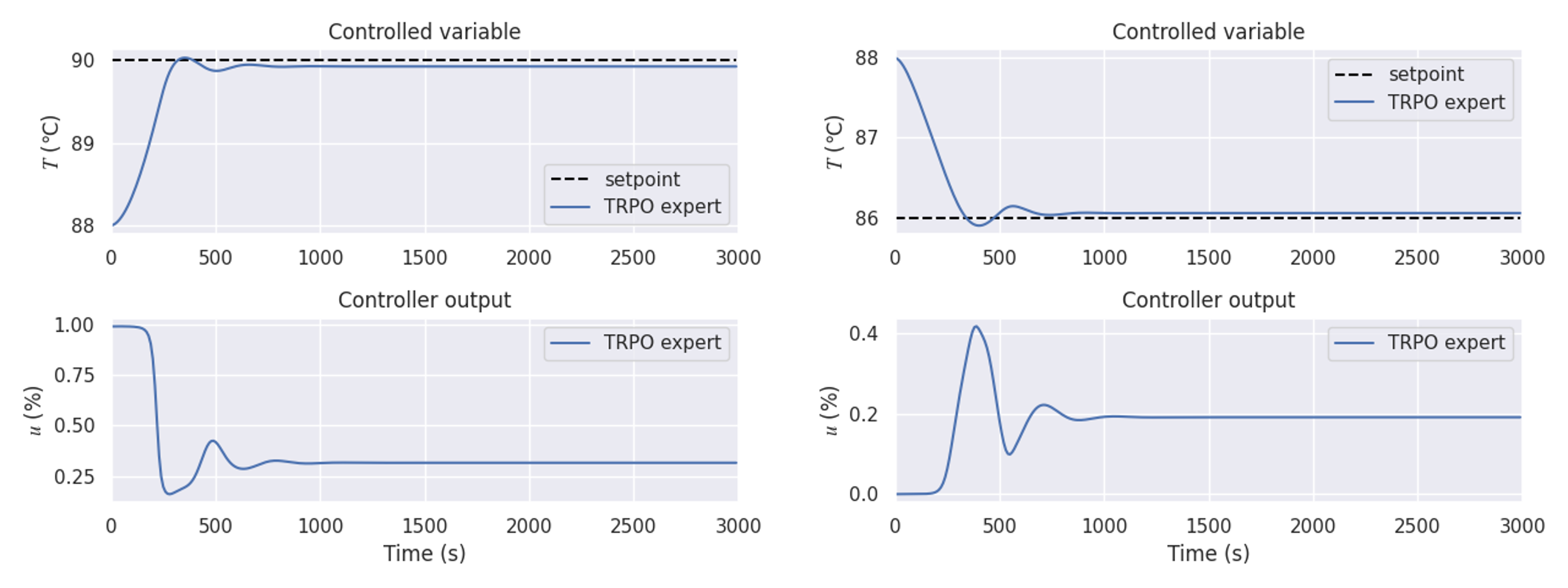}
	\caption{Typical control performances of the TRPO expert demonstrations (left: Mode 1 ${T^{{\rm{set}}}} = 90$; right: Mode 2 ${T^{{\rm{set}}}} = 86$).}
	\label{fig:5.4}
\end{figure}

\begin{figure}
	\centering
	\includegraphics[width=1.0\linewidth]{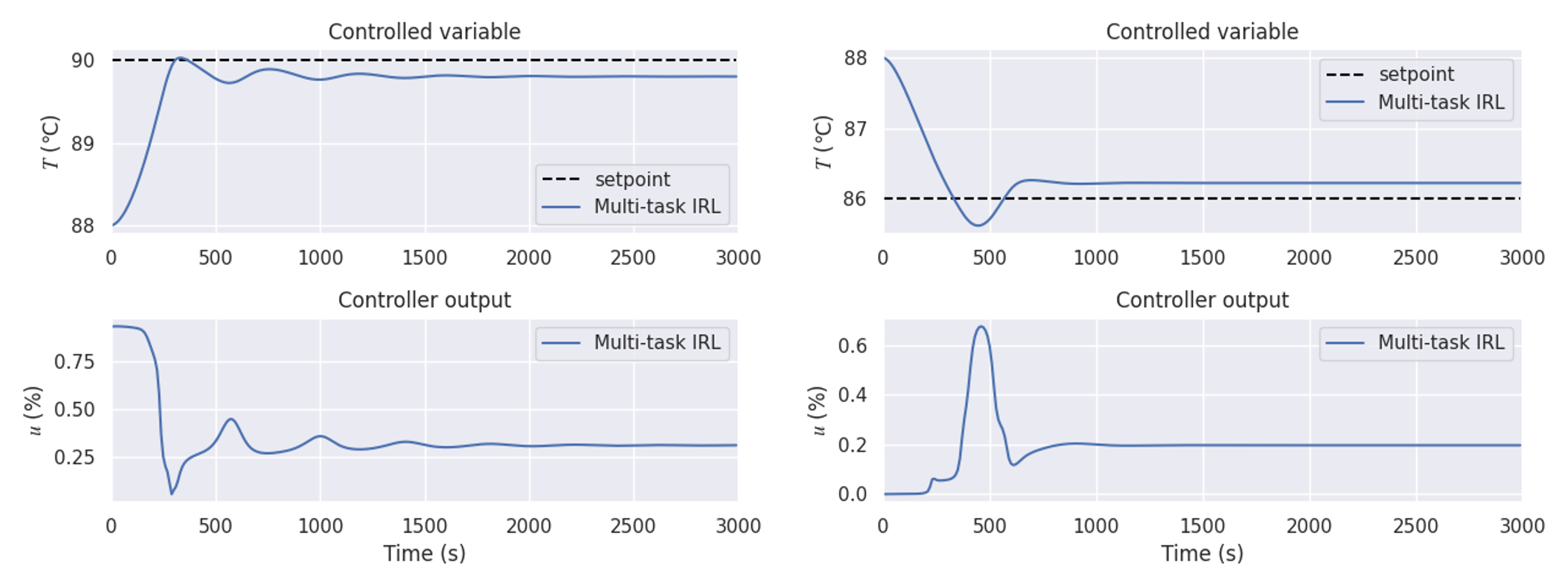}
	\caption{Control performances of the successfully trained multi-task IRL agent based on the TRPO expert demonstrations (left: Mode 1 ${T^{{\rm{set}}}} = 90$; right: Mode 2 ${T^{{\rm{set}}}} = 86$).}
	\label{fig:5.5}
\end{figure}

\noindent
As in Case 1, the TRPO algorithm and the reward function outlined in \cite{RN76, RN80} are employed to train the DRL-based expert. A total of 2,112 expert trajectories are collected across both modes. The typical TRPO expert trajectories for each mode are displayed in Fig. \ref{fig:5.4}, while the trajectories generated by the trained multi-task IRL agent are shown in Fig. \ref{fig:5.5}. These results are obtained by deploying the trained IRL controller directly into the environment for validation. While a small residual error remains after the controller stabilizes, the overall performance meets acceptable standards. Furthermore, in industrial applications, the pre-trained controller can be fine-tuned in real-world settings to facilitate Sim2Real transfer learning \cite{RN91, RN87}, potentially enhancing control performance further.

\subsubsection{Historical closed-loop operation data as expert trajectories}\noindent
To demonstrate the practical engineering potential and feasibility of using multi-task IRL for multi-mode process control, unknown expert demonstrations derived from historical closed-loop operating data will now serve as the training source. Compared to using RL agents as expert(s), adopting a classical control scheme (such as PID) as the expert controller offers a more compelling test. This is because the behavior patterns of RL agents and IRL agents are relatively similar, making the trajectory distribution generated by RL well-suited for IRL learning. In contrast, traditional PI control operates as a feedback control law based on error and cumulative error, presenting a notable challenge for IRL in accurately capturing its control characteristics.

\begin{figure}
	\centering
	\includegraphics[width=1.0\linewidth]{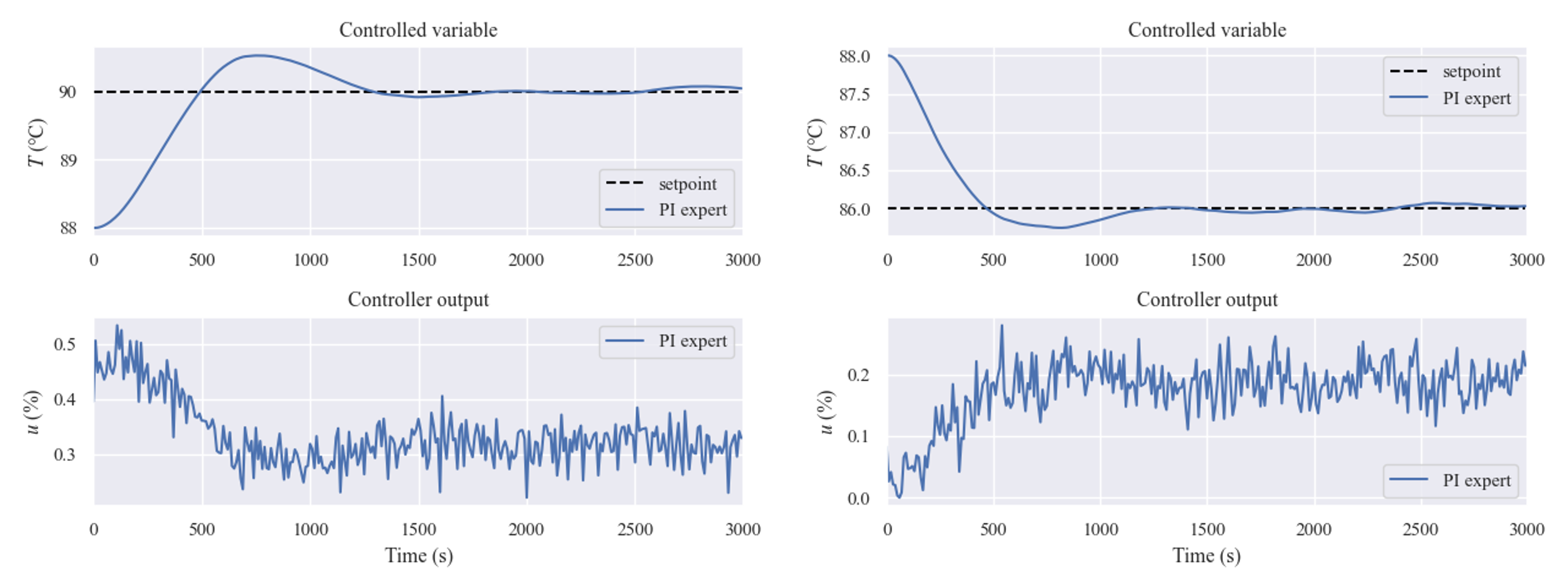}
	\caption{Typical control performances of the PI expert demonstrations (left: Mode 1 ${T^{{\rm{set}}}} = 90$; right: Mode 2 ${T^{{\rm{set}}}} = 86$).}
	\label{fig:5.6}
\end{figure}

In this experiment, a well-tuned PI controller for the CSTR system, is employed to generate closed-loop operating data for training the IRL agent. To introduce variability and stochasticity in the expert trajectories, white noise is applied to the inlet concentration. A total of 2,112 expert trajectories are recorded, each containing 300 samples (sampling interval $T_s$ = 10 seconds, corresponding to a total duration of 3,000 seconds for the system). The typical PI controller trajectories for the two modes are depicted in Fig. \ref{fig:5.6}, indicating that the PI controller maintains satisfactory control performance.

\begin{figure}
	\centering
	\includegraphics[width=1.0\linewidth]{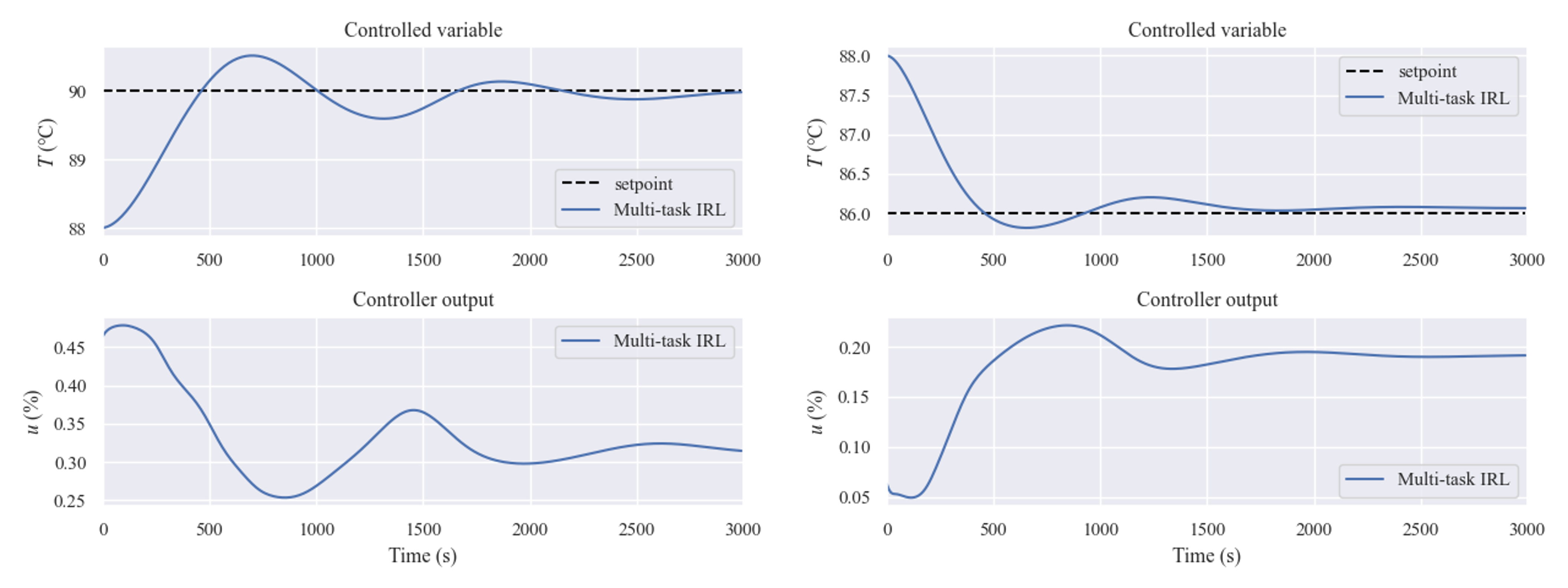}
	\caption{Control performances of the successfully trained multi-task IRL agent based on the PI expert demonstrations (left: Mode 1 ${T^{{\rm{set}}}} = 90$; right: Mode 2 ${T^{{\rm{set}}}} = 86$).}
	\label{fig:5.7}
\end{figure}

The closed-loop expert demonstrations described above are used for multi-task IRL training. The control performance of the trained multi-task IRL-based controller across the two modes is presented in Fig. \ref{fig:5.7}. The results indicate that the multi-task IRL agent effectively imitates expert behaviors and naturally adapts to varying modes. As in the previous case, if additional refinement is necessary, Sim2Real transfer learning can be employed to minimize model-plant mismatches.

\section{Conclusion}\label{Conclusion}
This paper presents a novel multi-task IRL approach aimed at addressing the multi-mode process control problem, with the primary objective of extracting controller patterns and RL value information from closed-loop big data encompassing multiple operating modes. To accomplish this, latent variables, commonly used in multi-mode modeling, are integrated to identify different mode indicators. Specifically, by introducing a latent context variable, the proposed method first establishes a mathematical framework to represent the conditional policy and trajectory distribution. Subsequently, techniques such as MaxEnt IRL, mutual information regularization, and variational inference are employed to optimize context-conditional rewards and policies. Experimental results demonstrate that the proposed method effectively learns a universal controller that can adapt to various scenarios based on multi-mode historical closed-loop data. This promising approach offers a probabilistic inference-based solution for data-driven controller design and underscores the potential of context-conditional latent variable modeling techniques in the development of multi-mode process controllers.



\bibliographystyle{IEEEtran}
\bibliography{ref}



 

\begin{IEEEbiography}[{\includegraphics[width=1in,height=1.25in,clip,keepaspectratio]{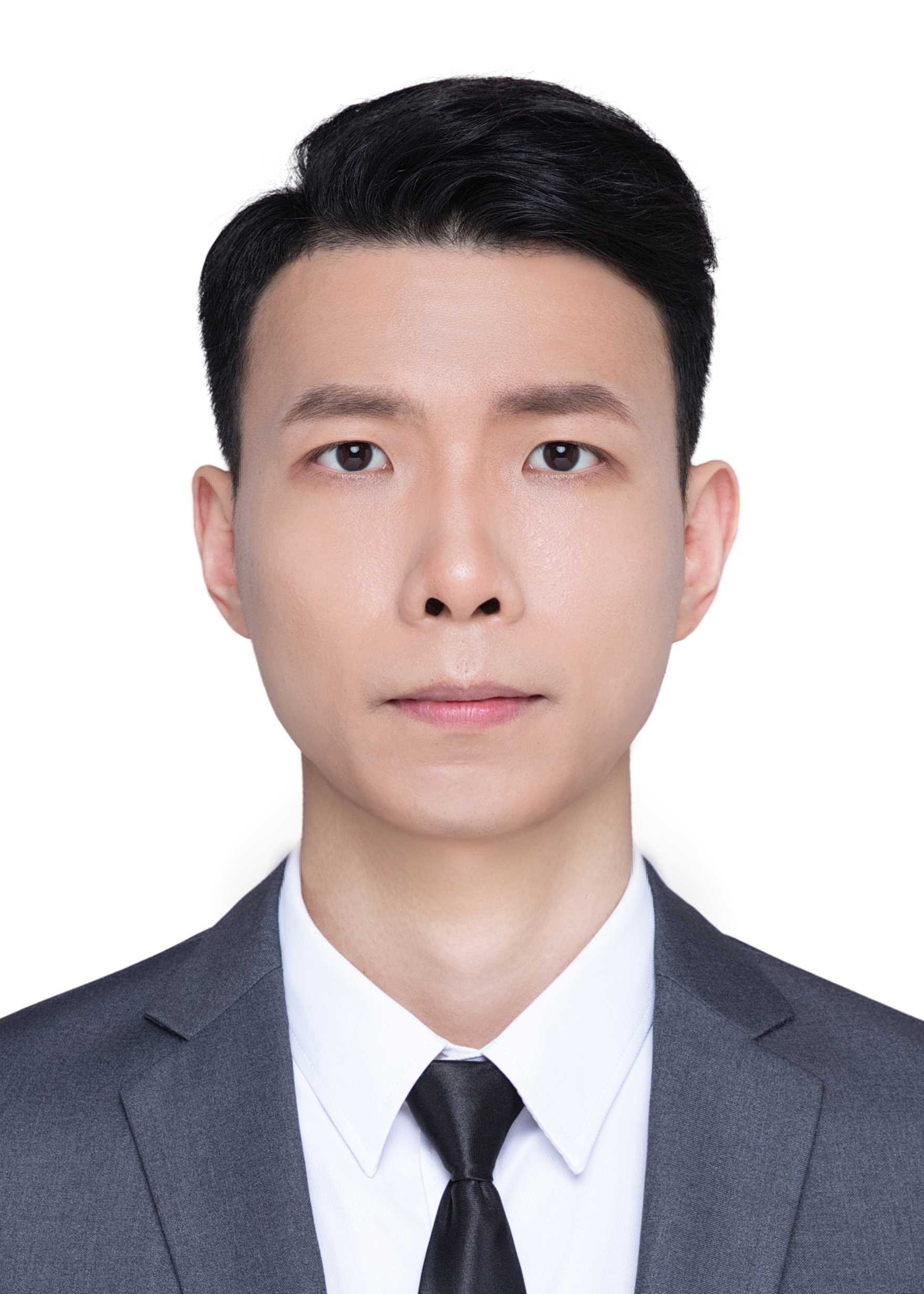}}]{Runze Lin}
received the B.S. degree in automation from the College of Control Science and Engineering, Zhejiang University, Hangzhou, China, in 2020, where he is currently pursuing the Ph.D. degree in control science and engineering with the State Key Laboratory of Industrial Control Technology, China.

From 2022 to 2023, he was a Visiting Scholar with the University of Alberta, Edmonton, AB, Canada. His research interests include reinforcement learning, transfer learning, process control, data analytics, industrial big data and its applications.
\end{IEEEbiography}

\vspace{-15pt}

\begin{IEEEbiography}[{\includegraphics[width=1in,height=1.25in,clip,keepaspectratio]{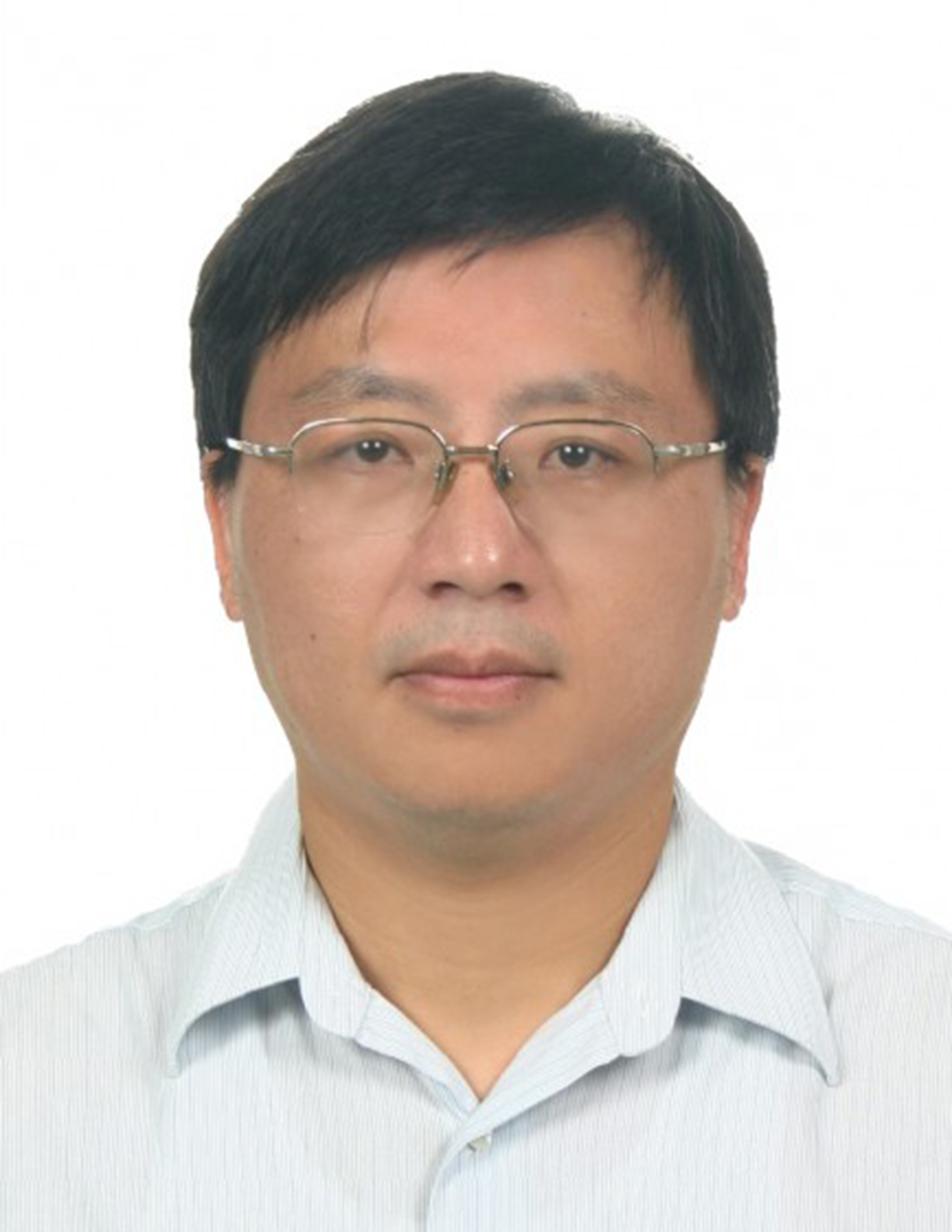}}]{Junghui Chen}
received the B.S. degree from the Department of Chemical Engineering, Chung Yuan Christian University, Taoyuan, Taiwan, in 1982, the M.S. degree from the Department of Chemical Engineering, National Taiwan University, Taipei, Taiwan, in 1984, and the Ph.D. degree from the Department of Chemical Engineering, The University of Tennessee at Knoxville, Knoxville, TN, USA, in 1995. 

He is currently a Full Professor with Chung Yuan Christian University. His research interests are process system engineering, including process design for operability, nonlinear control, process monitoring and diagnosis, control loop performance assessment, batch control, model predictive control, data mining and analytics, and iterative learning design.
\end{IEEEbiography}

\vspace{-15pt}

\begin{IEEEbiography}[{\includegraphics[width=1in,height=1.25in,clip,keepaspectratio]{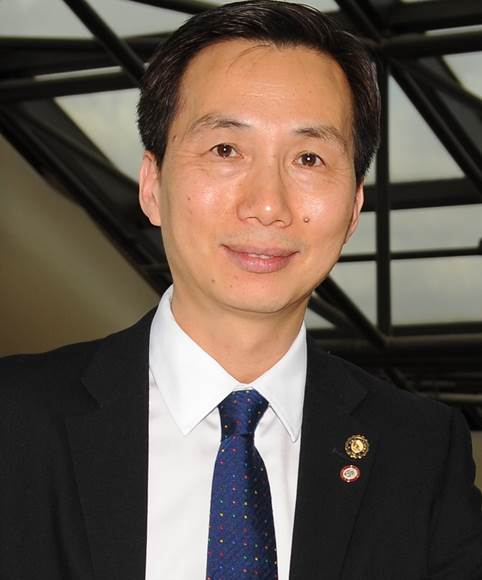}}]{Biao Huang}
(Fellow, IEEE) received the B.S. and M.S. degrees in automatic control from the Beijing University of Aeronautics and Astronautics, Beijing, China, in 1983 and 1986, respectively, and the Ph.D. degree in process control from the University of Alberta, Edmonton, AB, Canada, in 1997.

He joined the University of Alberta in 1997 as an Assistant Professor with the Department of Chemical and Materials Engineering, where he is currently a Full Professor. He was an NSERC Industrial Research Chair in control of oil sands processes and the AITF Industry Chair in process control from 2013 to 2018. His research interests include data analytics, process control, system identification, control performance assessment, Bayesian methods, and state estimation. He has applied his expertise extensively in industrial practice.

\end{IEEEbiography}

\vspace{-15pt}

\begin{IEEEbiography}[{\includegraphics[width=1in,height=1.25in,clip,keepaspectratio]{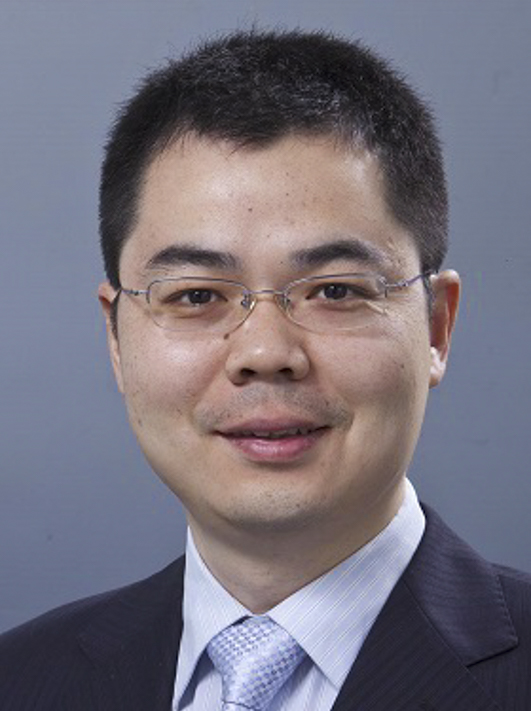}}]{Lei Xie}
received the B.S. and Ph.D. degrees from Zhejiang University, China, in 2000 and 2005, respectively.

From 2005 to 2006, he was a Postdoctoral Researcher with the Berlin University of Technology and an Assistant Professor from 2005 to 2008. He is currently a Professor with the Department of Control Science and Engineering, Zhejiang University. His research interests focus on the interdisciplinary area of statistics and system control theory.
\end{IEEEbiography}

\vspace{-15pt}

\begin{IEEEbiography}[{\includegraphics[width=1in,height=1.25in,clip,keepaspectratio]{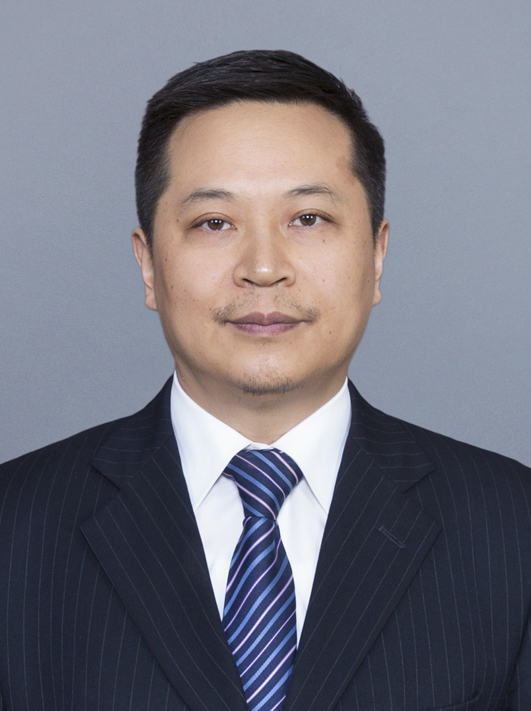}}]{Hongye Su}
(Senior Member, IEEE) received the B.S. degree in industrial automation from the Nanjing University of Chemical Technology, Jiangsu, China, in 1990, and the M.S. and Ph.D. degrees in industrial automation from Zhejiang University, Hangzhou, China, in 1993 and 1995, respectively.

From 1995 to 1997, he was a Lecturer with the Department of Chemical Engineering, Zhejiang University. From 1998 to 2000, he was an Associate Professor with the Institute of Advanced Process Control, Zhejiang University, where he is currently a Professor with the Institute of Cyber-Systems and Control. His current research interests include robust control, time-delay systems, and advanced process control theory and applications.
\end{IEEEbiography}

\vfill

\end{document}